\documentclass[twocolumn,showpacs,preprintnumbers,amsmath,amssymb]{revtex4}
\usepackage{graphicx}% Include figure files
\usepackage{dcolumn}% Align table columns on decimal point
\usepackage{bm}% bold math

\begin{document}

\preprint{}

\title{Living polymers in a size-asymmetric electrolyte}

\author{Sorin Bastea}
\email{bastea2@llnl.gov}
\affiliation{Lawrence Livermore National Laboratory, P.O. BOX 808, Livermore, CA 94550}

\date{\today}

\begin{abstract}
A living polymers transition is found in molecular dynamics simulations of a charge-symmetric 
size-asymmetric electrolyte with no anisotropic interactions. 
The fluid has strong polymeric character at low temperatures, where it consists 
of large, alternating-charge linear chains and rings in chemical equilibrium. A mean-field 
theory of chain association is consistent with the existence of such a transition.
In the polymeric phase the system is very weakly conducting or electrically insulating. 
\end{abstract}


\maketitle

Ionic systems are rather ubiquitous in nature, ranging from molten salts 
to the electrolyte solutions crucial for the functioning of living cells, 
to even pure water at very high densities \cite{ccstbp99}. A decisive step 
in understanding how the long-range character of the Coulomb potential 
determines the behavior of ionic systems has long proved to be the Debye-H\"uckel 
theory of screening by counterions \cite{dh23}. The 
introduction of the physically appealing notion of ionic pairing led to 
an additional refinement of these ideas and has allowed the recent development of 
successful theories for criticality in simple electrolyte models \cite{fl93}. The 
formation of ionic pairs and higher order clusters was initially 
observed and characterized in symmetric electrolytes \cite{mjg83} and it 
is now understood to be an essential feature of ionic systems \cite{lf96,ropf00,yp01,yp01jcp,wl01}.
Recent work has revealed that ionic clustering is even more pronounced in size-asymmetric model 
electrolytes and it is associated with an unexpected behavior of 
their phase coexistence regions \cite{ropf00,yp01}. Size-asymmetry is of course the norm 
in numerous circumstances, e.g. colloid science \cite{rh97} or systems with 
significant ionic character such as LiCl, NaH etc. We present molecular 
dynamics simulations of a size-asymmetric electrolyte with isotropic interactions, 
where strong ionic clustering produces a continuous polymerization transition to a phase 
with very weakly conducting or insulating character.  

We study a system of $N/2$ positive charges $+q$ and $N/2$ negative charges 
$-q$ interacting through potentials
\begin{eqnarray}
v_{ij}(r)=u_{ij}(r)+\frac{q_i q_j}{4\pi 
D_0 r}
\label{eq:one}.
\end{eqnarray}
($i,j=+,-$). $D_0$ is the vacuum dielectric permeability, and $u_{ij}(r)$ 
are \textit{exponential-6} potentials, 
\begin{eqnarray}
u_{ij}(r) = \epsilon\left[Ae^{-\alpha\frac{r}{r_{ij}}}-B\left(\frac{r_{ij}}{r}\right)^6\right]
\label{eq:two}
\end{eqnarray}
with well-depth $\epsilon$ corresponding to distance $r_{ij}$; $A=6e^{\alpha}/(\alpha-6)$, 
$B=\alpha/(\alpha-6)$ and $\alpha=13$. The Born-Mayer exponential 
term provides a good description of the repulsive short range interactions 
that occur between closed-shell atoms or ions, e.g. in molten salts \cite{gg94}. 
When modified to include many-body effects such simple ionic models appear 
to have a rather wide range of applicability \cite{mw00}.
We set $q=1$ in units of the electron charge $\mathit{e}$
and impose an extreme size-asymmetry by choosing $r_{++}=0$ and $r_{+-}=r_{--}/2$, 
i.e. the positive ions are point charges. We choose as relevant length scale the 
position of the minimum of $v_{+-}(r)$, which we denote by $d_{+-}$, $d_{+-}\simeq 0.67 r_{+-}$. 
This is appropriate at low temperatures where it should allow a comparison with the charged 
hard core models usually studied. The total reduced number 
density is $\rho^*=\rho d_{+-}^3$, $\rho=N/V$, and the reduced temperature 
$T^*=k_BT/E_0$, with $E_0=q^2/4\pi D_0d_{+-}$. 
Most of the simulations were performed with $N=2048$ particles in the microcanonical 
(\textit{NVE}) ensemble, in a box with periodic boundary conditions, 
at reduced density $\rho^*=0.12$ and reduced temperatures 
$T^*$ in the range $0.0065$ to $0.0365$. The reduced simulation box length was 
$L^*=L/d_{+-}\simeq 26$; simulations with $L^*\simeq 35$ ($N=5324$) were also 
performed to verify the influence of finite size effects. The energy scale of the short 
range potentials was set to $\epsilon/E_0\simeq 2.4\times 10^{-3}$. The Coulomb interactions 
were handled using the Ewald summation technique with conducting boundary conditions. 
The system was typically equilibrated for $5\times 10^4-10^5$ time steps, using 
successively velocity rescaling, Andersen thermostat and normal (\textit{NVE}) 
molecular dynamics. The accumulation runs consisted of at least $5\times 10^5$ time 
steps.

At the lowest temperatures the pair correlation functions signal a strongly structured
fluid, Fig. 1. The opposite-charge pair correlation function $g_{+-}(r)$ 
exhibits a very sharp peak at small distances, one that might be expected for example 
if significant ionic pairing takes place. An analysis of the coordination 
number as a function of distance, 
$N_{+-}(r) = (\rho/2)\int^{r}_{0} g_{+-}(r\prime) 4\pi {r\prime}^2 dr\prime$, 
Fig. 1 (inset), reveals however the presence of two opposite-charge nearest neighbors, an 
indication of alternating-charge chain association. The importance of such polymeric structures 
for the thermodynamics of electrolytes has been suggested before in the context of size-symmetric 
systems \cite{lf96} and has been recently reinforced by Monte Carlo simulations of 
size-asymmetric hard-core electrolytes \cite{ropf00,yp01}. Due to the large size-asymmetry their 
contribution is dominant here, apparently leading - see Fig. 1 - to a continuous transition 
between a non-structured isotropic fluid at high temperatures to a structured isotropic 
fluid at low temperatures. 
\begin{figure}
\includegraphics[width=8.0cm]{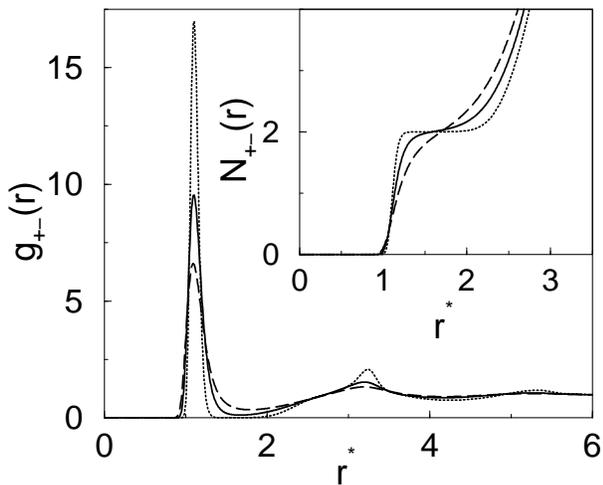}
\caption{\label{fig:fig1}Opposite-charge pair correlation functions $g_{+-}(r)$ at $T^*=0.0065$ (dotted line), 
$T^*=0.016$ (solid line), $T^*=0.024$ (dashed line) and corresponding coordination 
numbers $N_{+-}(r)$ (inset).}
\end{figure}

The formation of large structures in ionic systems can be understood using 
appropriate clustering definitions as first introduced for ionic systems by Gillan \cite{mjg83}.
Gillan's original analysis of cluster populations labels an ion as 
belonging to a cluster if the smallest distance between the ion and other cluster 
ions is less than a certain distance $R_C$. We use $R_C=1.5d_{+-}$, which is a good 
approximation for the position of the first minimum of $g_{+-}(r)$. We note however that the 
low temperatures analysis is insensitive to the choice of $R_C$ in a significant range. 
The same is true if we adopt a more restrictive definition of clustering, i.e. one 
that requires that an ion be 'linked' to a cluster through an ion of opposite charge.
We find that at low temperatures the system consists of large alternating-charge 
chains and rings that behave as living polymers, i.e. break and 
recombine continuously leading to a temperature-dependent equilibrium size distribution. 
The lifetime of these linear polymers, particularly the smaller ones, is rather long, 
which explains the extremely long times needed to equilibrate the system. 
\begin{figure}
\includegraphics[angle=-90, width=8.5cm]{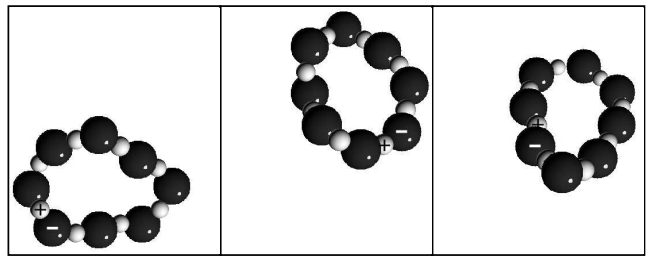}
\caption{\label{fig:fig2}Snapshots of the motion of a small alternating-charge ring at intervals of 
$ 150000$ time steps; $T^*=0.0065$.}
\end{figure}
\begin{figure}
\includegraphics[width=8.0cm]{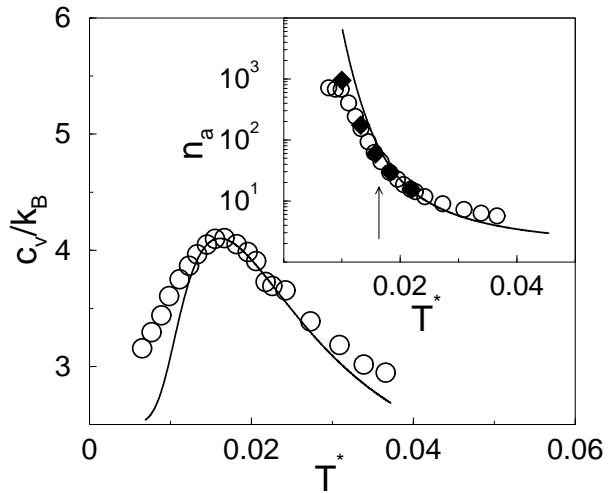}
\caption{\label{fig:fig3}Specific heat $c_V$ from simulations (circles) and mean-field 
theory (solid line) after a temperature rescaling (see text).
Inset: average chain size $n_a$ from simulations with 
$N=2048$ particles (circles), $N=5324$ (filled diamonds) and mean-field theory (solid line);
arrow indicates $T^*_p$.}
\end{figure}
Fig. 2 shows snapshots of the motion of such a small ring at the lowest temperature simulated, 
which diffuses as a well defined entity and exhibits the vibrational and rotational motions 
specific of a regular polymer.  

The association of neutral monomers in linear polymers under conditions of chemical equilibrium 
leads to a continuous polymerization transition accompanied by a maximum of 
the specific heat signaling the onset of chain formation \cite{rm95}. The position of the maximum is 
a measure of the association, or 'bond' energy within a chain. Such a transition 
is also apparently taking place in the ionic size-asymmetric system that we are simulating.  
We therefore calculate the constant volume specific heat, $c_V=(\partial E/\partial T)_V/N$, 
as a function of temperature in the microcanonical ensemble using the equilibrium kinetic 
energy fluctuations , 
$\langle \mathcal{K}^2 \rangle - \langle \mathcal{K} \rangle^2= (3/2)Nk_B^2 T^2(1 - 3 k_B/2 c_V)$ \cite{lpv67}. 
The maximum of $c_V$, Fig. 3, places the continuous polymerization transition 
at $T^*_p\simeq 0.016$, in agreement with the position of energy $E=E(T)$ and pressure 
$p=p(T)$ inflection points.
It is remarkable that in the living polymers systems previously studied the association of monomers 
into chains is driven by strong anisotropic interactions, which many times are of a dipolar 
nature \cite{ott96}. While in the present case no such anisotropy is present, isotropic 
long range (Coulomb) interactions mediated by extreme size-asymmetry lead to similar 
characteristic behavior. This could be perhaps attributed to an enhanced tendency for 
dipolar pairing due to size-asymmetry. However, although the chain size distribution 
above the polymerization temperature is peaked at $n=2$ (dipolar pairs), the distribution 
of ion fractions making up identical size chains is not, with the value for pairs remaining 
very small, of the order $10^{-2}$. Nevertheless, energy considerations also 
suggest that the formation of chains is favored in systems with large size-asymmetry 
\cite{yp01,ropf00}. If the propensity for chain formation is assumed to prevail for 
such electrolytes, a mean-field theory for charged associating chains 
can be formulated, as in the case of neutral living polymers \cite{cc90}. At the simplest 
level such a theory (see below) does not include the size-asymmetry \textit{per se}, but only 
implicitly in the chain association assumption. The system is assumed to be a mixture of 
alternating-charge chains which are electrically neutral when containing equal numbers of 
positive and negative ions and with charge $\pm q$ otherwise. At low densities the free energy 
of such a system can be written as 
\begin{eqnarray}
f=\frac{F}{V k_B T}=&&-\frac{3}{2}\rho \ln T+\sum_{n=1}\rho_{2n}^0(\ln\rho_{2n}^0-1)\nonumber\\
&&+\sum_{n=0}\left[\rho_{2n+1}^+(\ln\rho_{2n+1}^+ - 1)+\rho_{2n+1}^-(\ln\rho_{2n+1}^- - 1)\right]
\nonumber\\&& - \sum_{n=1}\rho_n\ln K_n + f_{DH}(\rho_c)
\end{eqnarray}
where $\rho_{2n}^0$ and $\rho_{2n}^{\pm}$ are densities of uncharged and charged chains, 
respectively. $K_n$ is the partition function of an $n-$chain and 
$\rho_c$ is the total density of charged chains $\rho_c=\sum_{n=0}(\rho_{2n+1}^++\rho_{2n+1}^-)$.
The Debye-H\"uckel term
\begin{equation}
f_{DH}(\rho_c)=-\frac{1}{4\pi d^3}\left[\ln (1+\kappa d) - \kappa d + \frac{(\kappa d)^2}{2}\right]
\end{equation}
includes an effective hard core $d$, that we set to $d_{+-}$ independent of polymerization 
effects; $\kappa^2=q^2\rho_c/D_0 k_B T$. 
For the $n-$chain partition function we use $K_n=[K_2(T)]^{n-1}$, with $K_2(T)$ the dimer 
association constant in the low temperature Bjerrum approximation, 
$K_2(T) = 4\pi d_{+-}^3 T^* exp(1/T^*)$ \cite{fl93}, neglecting interactions 
beyond nearest neighbors along a chain. 
(Here we treat the short range potential $u_{+-}(r)$ as effectively hard core, but this 
should introduce only small errors at low temperatures.) We note that if only $+-$ 'dimerization' 
is allowed, the above theory reduces to the so-called Debye-H\"uckel-Bjerrum (DHBj) theory \cite{fl93}.
The energy of the system is the sum of kinetic, 'bond' and ionic terms, 
\begin{equation}
e=\frac{E}{V}=\frac{3}{2}k_B T + (\rho-\rho_t)e_2+e_{DH}(\rho_c)
\end{equation}
with $\rho_t$ the total chains density, $\rho_t=\sum_{n=1} \rho_n$, $e_2= -E_0 + k_B T$ the average 
$+-$ 'bond' energy and $e_{DH}$ the Debye-H\"uckel contribution, 
\begin{equation}
e_{DH}(\rho_c)=-\frac{E_0 T^*}{8\pi d_{+-}^3}\frac{(\kappa d_{+-})^3}{1+\kappa d_{+-}}
\end{equation}
The chain size distribution $\rho_n$ is then obtained by minimizing $f$ at fixed density $\rho$ 
and temperature $T$, with the constraint of particle conservation, $\sum_{n=1} n\rho_n=\rho$, and 
charge neutrality, $\sum_{n=0}\left[\rho_{2n+1}^+-\rho_{2n+1}^-\right]=0$. We obtain
\begin{eqnarray} 
&&\rho_{2n}^0=K_2^{-1}e^{2n\lambda}\\
&&\rho_{2n+1}^+=\rho_{2n+1}^-=K_2^{-1}e^{(2n+1)\lambda}
exp\left(-\frac{\partial f_{DH}}{\partial \rho_c}\right)
\end{eqnarray}
where $\rho_{2n+1}^+=\rho_{2n+1}^-$ is a consequence of charge neutrality and $\lambda$ 
can be interpreted as the monomer chemical potential. The particle conservation 
condition can be recast as an equation for the charge density $\rho_c$, 
which after being solved numerically also allows the calculation of $\rho_t$. Substitution 
into the energy equation yields the specific heat through 
$\rho c_V=\partial e/\partial T|_\rho$. 

We find that $c_V$ displays the characteristic maximum \cite{rm95} of a 
living polymers transition at $T_{p0}^*\simeq 0.11$. 
This is higher than the transition temperature obtained in the simulations, but it is not 
surprising given the many approximations involved. In particular, the $K_n$ approximation used 
leads to a large overestimate even for the energy of long, straight alternating-charge chains, while 
size-asymmetry dependent packing and solvation effects, neglected here, should decrease 
the polymerization temperature by weakening the $+-$ 'bonds' \cite{lowdens}. After a suitable 
rescaling of the temperature the calculated $c_V$ is qualitatively compared with the simulation 
results in Fig. 3. We note that the height of the $c_V$ maximum is very well reproduced, 
along with the higher temperatures behavior. The low temperature deviations however are 
significant, probably due to the neglect of ring structures that appear to be very important below 
the transition temperature. In fact the cluster analysis yields mostly neutral structures at 
low temperatures, which is not well reproduced by the above mean-field model.

Another signature of a continuous polymerization transition is the rapid chain growth 
below $T^*_p$ \cite{rm95}. We show in Fig. 3(inset) the average chain size 
$n_a=\sum_{n=1}n\rho_n/\sum_{n=1}\rho_n$ found in the simulations and with the above 
mean-field treatment after the same rescaling of the temperature used for $c_V$. 
The chain size at $T^*_p$ is in 
good agreement with the theory, as it is the overall behavior. The slower rise of $n_a$ in 
the simulations than in the theory at low temperatures may be due to a depletion of linear 
chains due to rings formation, as noted before. It is worth pointing out that the finite size 
effects that appear to affect $n_a$ for the smaller system size occur only at the lowest 
temperatures, far away from $T^*_p$. We also estimate an effective  lifetime $\tau$ 
of the polymeric chains by assuming an initial exponential decay $exp(-t/\tau)$ of the time 
correlation function $C(\Delta t)=\langle \rho_t(t_0)\rho_t(t_0+\Delta t)\rangle - \langle \rho_t \rangle^2$ 
of the total number of chains $\rho_t$; $\tau$ is for example of the order of $20$ MD time 
steps at $T^*=0.01$ (compared with $\simeq 5\times 10^5$ total simulation steps), although 
it may be significantly longer for the smaller chains and rings.

The formation of large chains and rings and the dominance of the neutral ones 
has important consequences for the conduction properties of the polymeric phase.
We calculate the electrical conductivity $\sigma$ using the Green-Kubo expression
\begin{eqnarray}
\sigma(t)=\frac{1}{3Vk_B T}\int_0^t \langle \mathbf{j}(t\prime)\cdot\mathbf{j}(0)\rangle 
dt\prime
\end{eqnarray}
where $\mathbf{j}(t)=\sum_{k=1}^N q_k \mathbf{v}_k(t)$ is the charge current; $\sigma$ is 
evaluated from the large times value of $\sigma(t)$, $\sigma=\lim_{t\to\infty} \sigma(t)$. 
Below the transition temperature $\sigma(t)$ exhibits very strong oscillatory behavior and 
decays apparently to zero, i.e. the polymeric phase is very weakly conducting or insulating; 
the conductivities above $T_p^*$ are shown in Fig. 4. As the temperature increases the 
conductivity also increases exhibiting an 
\begin{figure}
\includegraphics[width=8.0cm]{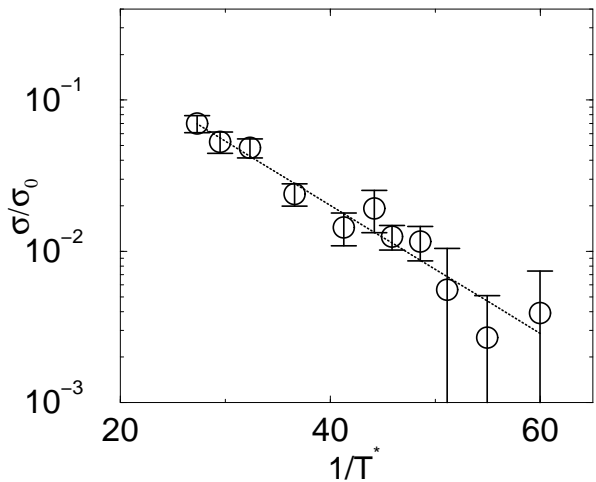}
\caption{\label{fig:fig5}Electrical conductivity for $T^*>T_p^*$ (circles) and 
fit with $\sigma=\sigma_0exp(-E_b^*/T^*)$ (dotted line).}
\end{figure}
activated behavior well described by 
$\sigma=\sigma_0exp(-E_b^*/T^*)$, with $E_b^*\simeq 0.1$. The conductivity increase is 
driven by an increasing charged clusters density, which occurs through the break-up of $+-$ 'bonds'. 
$E_b$ can therefore be interpreted as an effective 'bond' strength, yielding $T_p^*/E_b^*\simeq 0.16$, 
which is comparable with the value found in neutral living polymers of $0.25$ \cite{rm95}.
The possibility of an electrolyte insulating phase has been discussed before in the light of 
strong ionic pairing at low densities, in the vapor region of 
symmetric ionic systems \cite{wws00,jmc95,gg94}. An insulating phase may arise here in 
connection with a living polymers transition in the liquid region of the phase diagram.
It should be noted that the very slow dynamics of the system at low temperatures, leading 
to the observed Arrhenius temperature dependence of the conductivity, may render the question 
of the conducting or insulating character of the polymeric phase rather moot. 

The chain associating tendency of size-asymmetric electrolytes, underscored by the continuous 
polymerization transition reported here, may interfere with the vapor-liquid coexistence and 
perhaps explain the observed reduction of the vapor-liquid phase separation region in 
electrolytes as the size-asymmetry increases \cite{ropf00,yp01}. 
The competition between chain association and liquid condensation has been studied before 
for the case of neutral monomers, where it was 
found that vapor-liquid coexistence can become metastable under strong chain association 
conditions \cite{rr96}. This is also worth exploring in the case of ionic systems, particularly 
because simulation results suggest that vapor-liquid coexistence may persist in charge-symmetric 
size-asymmetric electrolytes even for the extreme case of point cations \cite{ropf00}, while theoretical 
calculations find no such coexistence \cite{rh97}. Finally, we note that polymeric behavior has 
been observed in molten $BeCl_2$ \cite{wm97}, where it was found to be due to the strong polarizability 
of the anion, coupled with the size-asymmetry of the ions. We find here that size-asymmetry alone 
may lead to living polymers behavior. 

I would like to thank Francis Ree and Joel Lebowitz for useful discussions. This work was performed 
under the auspices of the U. S. Department of Energy by University of California Lawrence 
Livermore National Laboratory under Contract No. W-7405-Eng-48.


\begin{thebibliography}{21}
\expandafter\ifx\csname natexlab\endcsname\relax\def\natexlab#1{#1}\fi
\expandafter\ifx\csname bibnamefont\endcsname\relax
  \def\bibnamefont#1{#1}\fi
\expandafter\ifx\csname bibfnamefont\endcsname\relax
  \def\bibfnamefont#1{#1}\fi
\expandafter\ifx\csname citenamefont\endcsname\relax
  \def\citenamefont#1{#1}\fi
\expandafter\ifx\csname url\endcsname\relax
  \def\url#1{\texttt{#1}}\fi
\expandafter\ifx\csname urlprefix\endcsname\relax\def\urlprefix{URL }\fi
\providecommand{\bibinfo}[2]{#2}
\providecommand{\eprint}[2][]{\url{#2}}

\bibitem[{\citenamefont{\text{C. Cavazzoni et al.}}(1999)}]{ccstbp99}
\bibinfo{author}{\bibnamefont{\text{C. Cavazzoni et al.}}},
  \bibinfo{journal}{Science} \textbf{\bibinfo{volume}{283}},
  \bibinfo{pages}{44} (\bibinfo{year}{1999}).

\bibitem[{\citenamefont{\text{P. W. Debye and E. H\"uckel}}(1923)}]{dh23}
\bibinfo{author}{\bibnamefont{\text{P. W. Debye and E. H\"uckel}}},
  \bibinfo{journal}{Phys. Z.} \textbf{\bibinfo{volume}{24}},
  \bibinfo{pages}{185} (\bibinfo{year}{1923}).

\bibitem[{\citenamefont{Fisher and Levin}(1993)}]{fl93}
\bibinfo{author}{\bibfnamefont{M.~E.} \bibnamefont{Fisher}} \bibnamefont{and}
  \bibinfo{author}{\bibfnamefont{Y.}~\bibnamefont{Levin}},
  \bibinfo{journal}{Phys. Rev. Lett.} \textbf{\bibinfo{volume}{71}},
  \bibinfo{pages}{3826} (\bibinfo{year}{1993}).

\bibitem[{\citenamefont{Gillan}(1983)}]{mjg83}
\bibinfo{author}{\bibfnamefont{M.~J.} \bibnamefont{Gillan}},
  \bibinfo{journal}{Molecular Physics} \textbf{\bibinfo{volume}{49}},
  \bibinfo{pages}{421} (\bibinfo{year}{1983}).

\bibitem[{\citenamefont{Levin and Fisher}(1996)}]{lf96}
\bibinfo{author}{\bibfnamefont{Y.}~\bibnamefont{Levin}} \bibnamefont{and}
  \bibinfo{author}{\bibfnamefont{M.~E.} \bibnamefont{Fisher}},
  \bibinfo{journal}{Physica A} \textbf{\bibinfo{volume}{225}},
  \bibinfo{pages}{164} (\bibinfo{year}{1996}).

\bibitem[{\citenamefont{Romero-Enrique
  et~al.}(2000)\citenamefont{Romero-Enrique, Orkoulas, Panagiotopoulos, and
  Fisher}}]{ropf00}
\bibinfo{author}{\bibfnamefont{J.~M.} \bibnamefont{Romero-Enrique}},
  \bibinfo{author}{\bibfnamefont{G.}~\bibnamefont{Orkoulas}},
  \bibinfo{author}{\bibfnamefont{A.~Z.} \bibnamefont{Panagiotopoulos}},
  \bibnamefont{and} \bibinfo{author}{\bibfnamefont{M.~E.}
  \bibnamefont{Fisher}}, \bibinfo{journal}{Phys. Rev. Lett.}
  \textbf{\bibinfo{volume}{85}}, \bibinfo{pages}{4558} (\bibinfo{year}{2000}).

\bibitem[{\citenamefont{Yan and de~Pablo}(2001{\natexlab{a}})}]{yp01}
\bibinfo{author}{\bibfnamefont{Q.}~\bibnamefont{Yan}} \bibnamefont{and}
  \bibinfo{author}{\bibfnamefont{J.~J.} \bibnamefont{de~Pablo}},
  \bibinfo{journal}{Phys. Rev. Lett.} \textbf{\bibinfo{volume}{86}},
  \bibinfo{pages}{2054} (\bibinfo{year}{2001}{\natexlab{a}}).

\bibitem[{\citenamefont{Yan and de~Pablo}(2001{\natexlab{b}})}]{yp01jcp}
\bibinfo{author}{\bibfnamefont{Q.}~\bibnamefont{Yan}} \bibnamefont{and}
  \bibinfo{author}{\bibfnamefont{J.~J.} \bibnamefont{de~Pablo}},
  \bibinfo{journal}{J. Chem. Phys.} \textbf{\bibinfo{volume}{114}},
  \bibinfo{pages}{1727} (\bibinfo{year}{2001}{\natexlab{b}}).

\bibitem[{\citenamefont{Weiss and Levesque}(2001)}]{wl01}
\bibinfo{author}{\bibfnamefont{J.~J.} \bibnamefont{Weiss}} \bibnamefont{and}
  \bibinfo{author}{\bibfnamefont{D.}~\bibnamefont{Levesque}},
  \bibinfo{journal}{Chem. Phys. Lett.} \textbf{\bibinfo{volume}{336}},
  \bibinfo{pages}{523} (\bibinfo{year}{2001}).

\bibitem[{\citenamefont{van Roij and Hansen}(1997)}]{rh97}
\bibinfo{author}{\bibfnamefont{R.}~\bibnamefont{van Roij}} \bibnamefont{and}
  \bibinfo{author}{\bibfnamefont{J.-P.} \bibnamefont{Hansen}},
  \bibinfo{journal}{Phys. Rev. Lett.} \textbf{\bibinfo{volume}{79}},
  \bibinfo{pages}{3082} (\bibinfo{year}{1997}).

\bibitem[{\citenamefont{Guissani and Guillot}(1994)}]{gg94}
\bibinfo{author}{\bibfnamefont{Y.}~\bibnamefont{Guissani}} \bibnamefont{and}
  \bibinfo{author}{\bibfnamefont{B.}~\bibnamefont{Guillot}},
  \bibinfo{journal}{J. Chem. Phys.} \textbf{\bibinfo{volume}{101}},
  \bibinfo{pages}{490} (\bibinfo{year}{1994}).

\bibitem[{\citenamefont{Madden and Wilson}(2000)}]{mw00}
\bibinfo{author}{\bibfnamefont{P.~A.} \bibnamefont{Madden}} \bibnamefont{and}
  \bibinfo{author}{\bibfnamefont{M.}~\bibnamefont{Wilson}},
  \bibinfo{journal}{J. Phys. Condens. Matter} \textbf{\bibinfo{volume}{12}},
  \bibinfo{pages}{A95} (\bibinfo{year}{2000}).

\bibitem[{\citenamefont{Rouault and Milchev}(1995)}]{rm95}
\bibinfo{author}{\bibfnamefont{Y.}~\bibnamefont{Rouault}} \bibnamefont{and}
  \bibinfo{author}{\bibfnamefont{A.}~\bibnamefont{Milchev}},
  \bibinfo{journal}{Phys. Rev. E} \textbf{\bibinfo{volume}{51}},
  \bibinfo{pages}{5905} (\bibinfo{year}{1995}).

\bibitem[{\citenamefont{Lebowitz et~al.}(1967)\citenamefont{Lebowitz, Percus,
  and Verlet}}]{lpv67}
\bibinfo{author}{\bibfnamefont{J.~L.} \bibnamefont{Lebowitz}},
  \bibinfo{author}{\bibfnamefont{J.~K.} \bibnamefont{Percus}},
  \bibnamefont{and} \bibinfo{author}{\bibfnamefont{L.}~\bibnamefont{Verlet}},
  \bibinfo{journal}{Phys. Rev.} \textbf{\bibinfo{volume}{153}},
  \bibinfo{pages}{250} (\bibinfo{year}{1967}).

\bibitem[{\citenamefont{Osipov et~al.}(1996)\citenamefont{Osipov, Teixeira, and
  Telo~da~Gama}}]{ott96}
\bibinfo{author}{\bibfnamefont{M.~A.} \bibnamefont{Osipov}},
  \bibinfo{author}{\bibfnamefont{P.~I.~C.} \bibnamefont{Teixeira}},
  \bibnamefont{and} \bibinfo{author}{\bibfnamefont{M.~M.}
  \bibnamefont{Telo~da~Gama}}, \bibinfo{journal}{Phys. Rev. E}
  \textbf{\bibinfo{volume}{54}}, \bibinfo{pages}{2597} (\bibinfo{year}{1996}).

\bibitem[{\citenamefont{Cates and Candau}(1990)}]{cc90}
\bibinfo{author}{\bibfnamefont{M.~E.} \bibnamefont{Cates}} \bibnamefont{and}
  \bibinfo{author}{\bibfnamefont{S.~J.} \bibnamefont{Candau}},
  \bibinfo{journal}{J. Phys. Condense. Matter} \textbf{\bibinfo{volume}{2}},
  \bibinfo{pages}{6869} (\bibinfo{year}{1990}).

\bibitem[{\citenamefont{\text{Preliminary simulations with $\rho^*=0.3$ yield
  $T_p^*\simeq 0.02$.}}()}]{lowdens}
\bibinfo{author}{\bibnamefont{\text{Preliminary simulations with $\rho^*=0.3$
  yield $T_p^*\simeq 0.02$.}}}

\bibitem[{\citenamefont{\text{H. Weinga\"rtner, V. C. Weiss, and W.
  Schro\"rer}}(2000)}]{wws00}
\bibinfo{author}{\bibnamefont{\text{H. Weinga\"rtner, V. C. Weiss, and W.
  Schro\"rer}}}, \bibinfo{journal}{J. Chem. Phys.}
  \textbf{\bibinfo{volume}{113}}, \bibinfo{pages}{762} (\bibinfo{year}{2000}).

\bibitem[{\citenamefont{Callilol}(1995)}]{jmc95}
\bibinfo{author}{\bibfnamefont{J.~M.} \bibnamefont{Callilol}},
  \bibinfo{journal}{J. Chem. Phys.} \textbf{\bibinfo{volume}{102}},
  \bibinfo{pages}{5471} (\bibinfo{year}{1995}).

\bibitem[{\citenamefont{van Roij}(1996)}]{rr96}
\bibinfo{author}{\bibfnamefont{R.}~\bibnamefont{van Roij}},
  \bibinfo{journal}{Phys. Rev. Lett.} \textbf{\bibinfo{volume}{76}},
  \bibinfo{pages}{3348} (\bibinfo{year}{1996}).

\bibitem[{\citenamefont{Wilson and Madden}(1997)}]{wm97}
\bibinfo{author}{\bibfnamefont{M.}~\bibnamefont{Wilson}} \bibnamefont{and}
  \bibinfo{author}{\bibfnamefont{P.~A.} \bibnamefont{Madden}},
  \bibinfo{journal}{Mol. Phys.} \textbf{\bibinfo{volume}{92}},
  \bibinfo{pages}{197} (\bibinfo{year}{1997}).

\end{thebibliography}
\end{document}